# The spin glass-like dynamics of gelatin gels[*]

Alan Parker[°] and Valéry Normand

Corporate R&D Division, Firmenich SA, 7, rue de la Bergère, 1217 Meyrin 2 Geneva, Switzerland

[°] Corresponding author. E-mail: alan.parker@firmenich.com

Abstract

*We show that there are several striking parallels between the dynamics of gelatin gels and spin glasses. In general, glassy systems retain a memory of their past history. A key characteristic differentiating spin glasses from most other glassy systems is that on cooling they appear to "forget" what happened just below the glass transition temperature, but the memory is recovered on heating. We show that gelatin gels also behave in this way. Both systems show critical scaling of the kinetics with temperature and undergo physical aging, that is they never reach equilibrium, but continue to harden indefinitely at a rate which is linear in log(time). The parallels between the dynamics of these two completely different kinds of condensed matter strongly suggest that they share an underlying theory.*

There has recently been an explosion of interest in the physics of systems trapped far from equilibrium [1, 2]. Intriguing parallels have been found between the behavior of systems as diverse as traffic, powders, soft matter, spin glasses and structural glasses, such as window glass. A central issue is to define when these parallels are superficial and when they are profound. The answers are extremely hard to find. Thermo-reversible gels are clearly trapped far from equilibrium. In fact, more than twenty years ago de Gennes remarked that such weakly cross-linked gels "should show some of the intricacy of the glass transitions" [3]. This suggestion has never been tested. Here we show more precisely that several aspects of the dynamics of gelatin, the archetypal thermo-reversible gel, are astonishingly close to those of spin glasses.

Spin glasses have long served as a source of inspiration for modeling systems with many almost equivalent energy minima, for example memory storage in the brain [4] and optimization problems [5]. This appeal is due to a combination of the simplicity of the theoretical spin glass models and the richness of the resulting behavior. This behavior is still far from being completely understood [1], despite huge theoretical, numerical and experimental efforts.

Gelatin is degraded collagen. When its solutions are cooled below about 40°C, the separate chains start to combine and re-form portions of collagen triple helix, which cross link the system, eventually forming an elastic gel. Djabourov *et al.* measured the elasticity and helix fraction as a function of time after sudden cooling to a range of temperatures [6]. They showed that the sol-gel transition is a percolation threshold. Once chains are involved in at least two cross-links, the system becomes frustrated, due to the competition between neighboring cross-links for the shared portions of free chain. Frustration, the incompatibility of local and global energy minimization, is a key feature of all glassy systems [7]. Gelatin gels have a gelation temperature above which a gel will never form. The thermodynamic status of this temperature has never been clarified. We show that it shares some characteristics with the critical temperature of a second order phase transition.

Gelatin gels display two of the key features of glassy systems:
1) The mechanical response occurs at two widely separated time scales [8], which is typical of structural glasses (see, for instance, [9]).
2) For aging times between one hour and several months the elastic modulus of gelatin gels increases as log(time) [10]. For glassy systems in general, the rate of physical aging is proportional to $(\log(\text{time}))^\zeta$ with $\zeta$ close to 1 [1].

When a glassy system is cooled below the glass transition temperature, $T_g$, equilibrium is never reached [1]. The system properties then depend on the time spent below $T_g$. In general, glassy systems remember their past history. The key characteristic that distinguishes spin glasses from other glassy systems is that memory can be temporarily lost [11]. This effect is best observed using a two stage protocol: A non-perturbing alternating field is applied (magnetic for spin glasses, mechanical for gels) and the response (magnetic susceptibility for spin glasses, elasticity for gels) is measured whilst: 1) cooling at a constant rate from above $T_c$ to below it and then heating at the same rate. This gives the reference curve. 2) Cooling and heating in the same way, but stopping once or twice during the cooling. Figure 1 shows the results for experiments with one stop (1A and 1C) and two stops (1B and 1D). Figures 1A,B show the raw data and figures 1C,D show the difference between the experiments with and without stops ($\Delta G' = G' - G'_{ref}$).

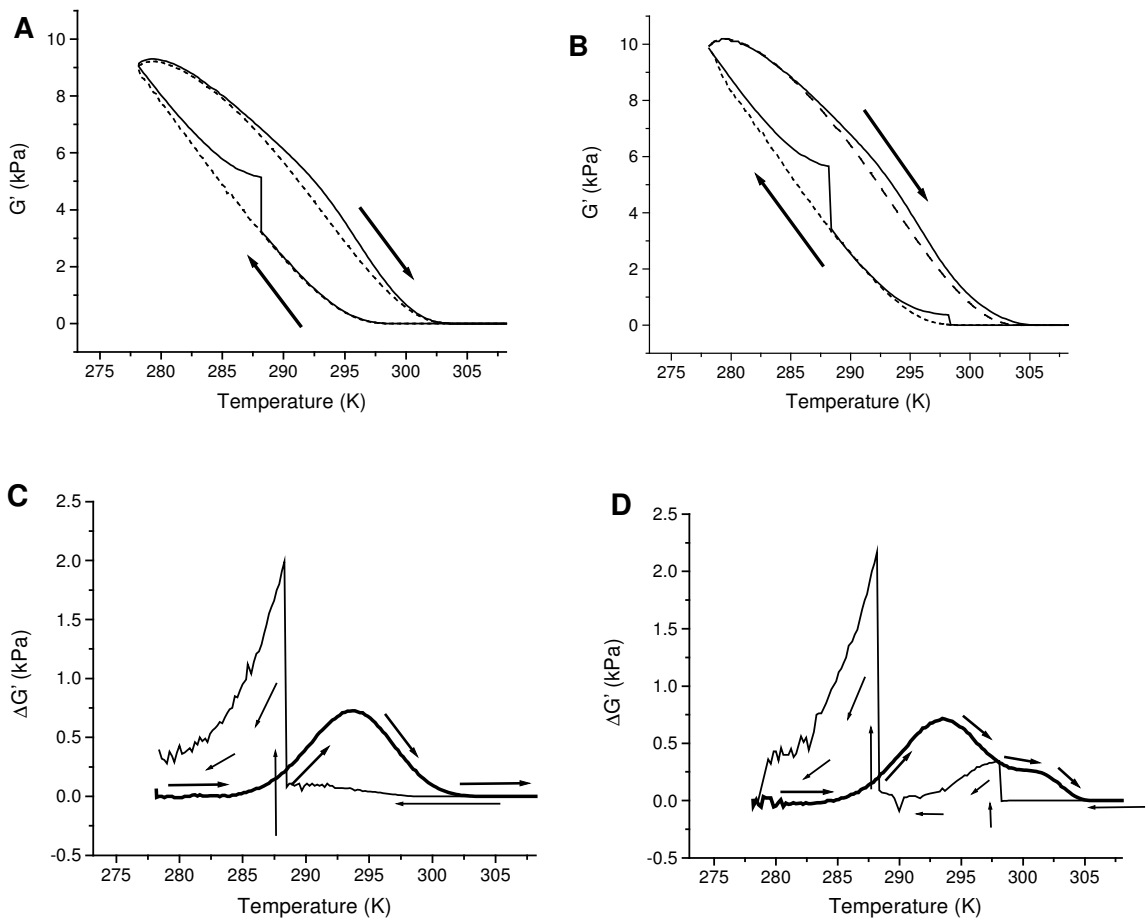

*Figure 1 The memory effect for gelatin gels. Top row shows elasticity measured during cooling and then heating at 0.2Kmin$^{-1}$. The dashed line is the reference curve measured during continuous cooling and heating. The solid line is the result with stopping. Bottom row shows the difference between the reference curve and the stopping curve. The thin line is for cooling and the bold line for heating. On the left one stop: 1h at 15°C. On the right two stops: 2h at 25°C and 1h at 15°C. c = 100g/kg*

These results are very similar to those for spin glasses [11] and also for a polymer glass [12] in that: i) at low temperature the sample shows the same behavior with or without stopping, it temporarily forgets its past, ii) on heating the sample remembers that it stopped, which causes a melting peak to appear close to the stopping temperature(s). A further similarity with spin glasses [13] is that the response after two stops is just the sum of the responses to two separate stops. For spin glasses the "melting" peak occurs exactly at the stopping temperature. For our gels (and Bellon et al.'s polymer glass [12]) the melting peak occurs 10°C higher. Figure 2 shows that this is an effect of heating rate.

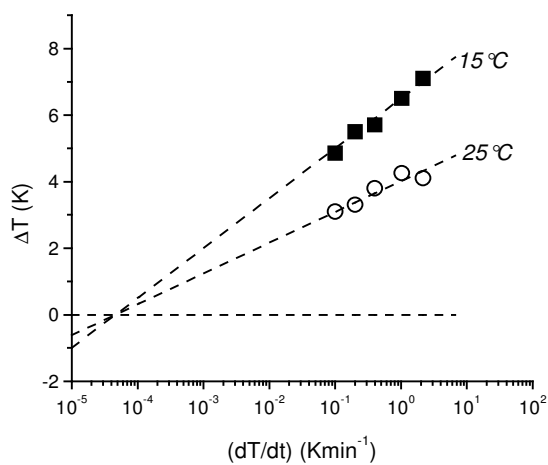

*Figure 2 Effect of heating rate on temperature difference between stopping temperature and temperature of melting peak for stopping temperatures of 15°C and 25°C. Extrapolations by eye.*

All the gels were cooled at the same rate, but heated at different rates. Figure 2 shows the temperature difference between the stopping temperature and the peak rate during melting as a function of the heating rate. Lower heating rates lead to smaller differences. The extrapolations suggest that the melting and stopping temperatures would coincide at a heating rate of less than $5 \times 10^{-5}$ Kmin$^{-1}$. We think that this difference between spin glasses and gelatin gels is just due to the much slower dynamics of the gels.

It was shown recently [10] that the gelation kinetics of a wide range of gelatin gels can be superposed on an arbitrary reference curve by shifting them in log(modulus)/log(time) space. The majority of the data could be fit by a linear dependence of the shift factors on temperature, concentration and molecular weight distribution. We call this regime far-from-critical. However, it was clear that a different scaling was required when the temperature was too high or the concentration too low. We call this regime close-to-critical. Figure 3 shows some typical data for the elastic modulus as a function of time in the close-to-critical regime. Note how strongly the kinetics depend on the temperature.

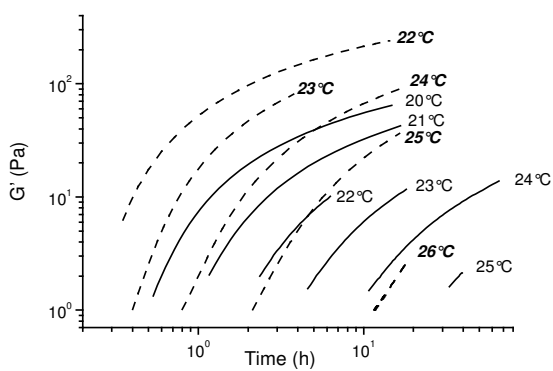

*Figure 3 Typical gelation kinetics in the close-to-critical regime. Concentrations: Solid lines: 20g/kg, Dashed lines: 40g/kg.*

We improve on the previous work [10] by showing that these data can be collapsed using the concept of critical slowing down [14]. Note that although critical slowing down is characteristic of second order phase transitions, it is not proof of the presence of a second order phase transition. It is just a "catastrophe flag" [14] indicating that the system is close to a singular point. The relaxation time, $\tau$, diverges as: $\tau \propto |X_c - X|^n$, with $X$ a system variable, $X_c$ its critical value and $n$ an exponent. Hukushima et al. [15] showed that this dynamic scaling accounted for the effect of temperature on the time evolution of their spin-glass simulations. We apply this approach to the gelation kinetics of gelatin. Interestingly, we find that it can be extended to account for the effect of concentration too. We fit the data in figure 3 to the formula:

$$\frac{G'(t)}{\varepsilon^{\alpha}(c-c_c)^{\mu}} = g\left(\frac{t}{\varepsilon^{\beta}(c-c_c)^{\nu}}\right) \quad ; \quad \varepsilon = 1 - \frac{T}{T_c} \qquad (1)$$

where $\varepsilon$ is the reduced temperature, $c$ the dimensionless concentration, $t$ is time and $g(x)$ is a scaling function defining the shape of the master-curve. The four exponents and the critical concentration, $c_c$, are fitting parameters. The form of equation (1) is evidence against the presence of the second order phase transition, as the effects of concentration and temperature are independent. Figure 4 shows that the best fit of equation (1) to the data in Figure 2 gives an excellent collapse onto a single curve.

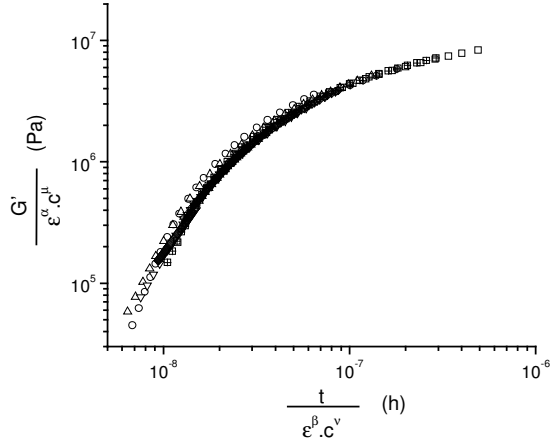

*Figure 4 Critical scaling of data in figure 2. Solid symbols 20g/kg, open symbols 40g/kg. Best fit values in equation (1) were: $\alpha$ = 3.23 ± 0.09, $\beta$ = -9.30 ± 0.13, $\mu$ = 2.3, $\nu$ = -2.6, $c_c$ = 0, $T_c$ = 35.8°C.*

Data for concentrations between 7.5 and 100g/kg also fell onto the same curve. Setting the critical concentration, $c_c$, to zero gave the best fit, implying that the critical gelation concentration is much lower than the lowest concentration studied. We argue that this approach is the only non-arbitrary way of determining the critical gelation concentration, as it is time-independent.

Figure 5 shows that Equation (1) only collapses the data in the far-from-critical regime for short times. At longer times, the gelation rate is slower.

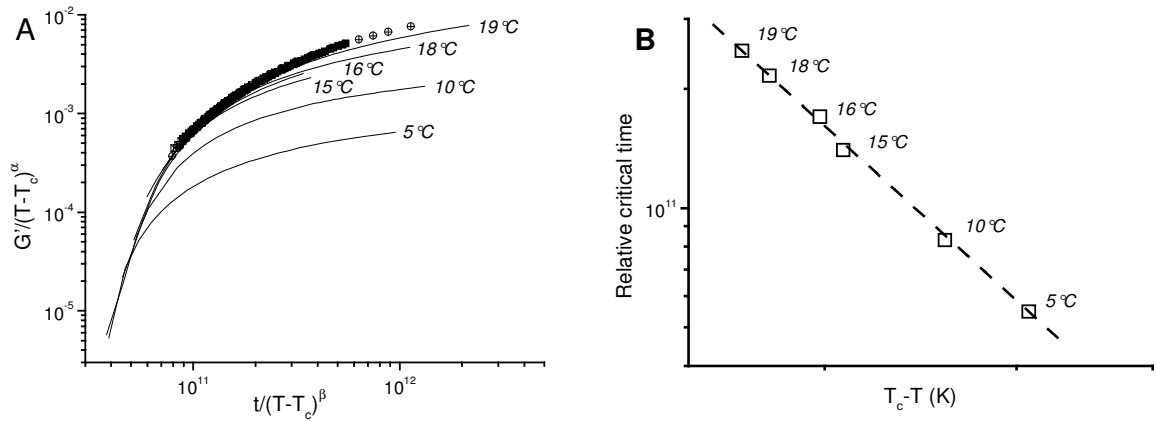

*Figure 5A Deviation of the gelation kinetics of cold-aged gels from the master curve for close-to-critical gels, shown in fig. 2. 5B shows that the time at which the deviation occurs scales with the distance from the critical temperature.*

In fact, Equation (1) can be used to fit data in the far from critical regime too (data not shown), using the same $T_c$ as for the close-to-critical regime, but different exponents. In particular, the time exponent of the reduced temperature, β, falls from about -9 to close to -2, so when gels are far from critical they evolve much more slowly than when they are close to it. Finally, figure 6 shows that the experimentally well defined time at which the transition between the two regimes occurs also scales with concentration and temperature.

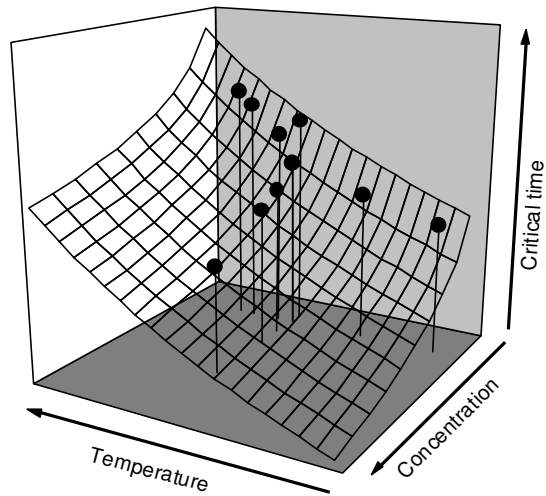

*Figure 6 Surface defining the critical time that separates the close-to-critical (below the surface) and far-from-critical regimes (above).*

Two regimes with different scaling exponents are also found for second order phase transitions [16]. Close-to-critical, the dynamics are dominated by fluctuations from the average behavior. Far-from-critical the dynamics are controlled by thermally activated hopping over energy barriers. Also the system evolves more slowly in the far-from-critical regime, as we have shown for gelatin gels. De Gennes suggested long ago that these ideas were relevant to gels [17], but their validity has never been established [18]. As we pointed out above, the presence of aging and the form of equation (1) both argue against the idea that the behavior of gelatin gels is related to a second order phase transition. Recent work on real spin glasses [19] suggests that their dynamics also show these two regimes. However, the relevance of the second order phase transition to the behavior of real spin glasses is still an open question [20]. For gelatin, direct measurement of the fluctuations, using light scattering or high sensitivity rheometry [21], would be very helpful.

Gelatin gels close to and far from critical are not only differentiated by their scaling exponents. There is also a correlation with a very simple observation that was reported previously [10]: a gel is cut in two and the pieces placed back together. A cut made when the gel is close to critical will heal, whereas a cut made when it is far from critical will not. Intuitively, this observation agrees with the fact that gels in the far-from-critical regime show more dramatic memory effects, as we show below.

The memory effect has been studied in real [19] and simulated [22] spin glasses after sudden heating of an aged sample to a temperature below $T_c$. Figure 4A shows that an equivalent memory effect occurs in the close-to-critical regime for gelatin. Gels were first aged at 15°C for up to 1 hour and then heated to 24°C, well below the critical temperature.

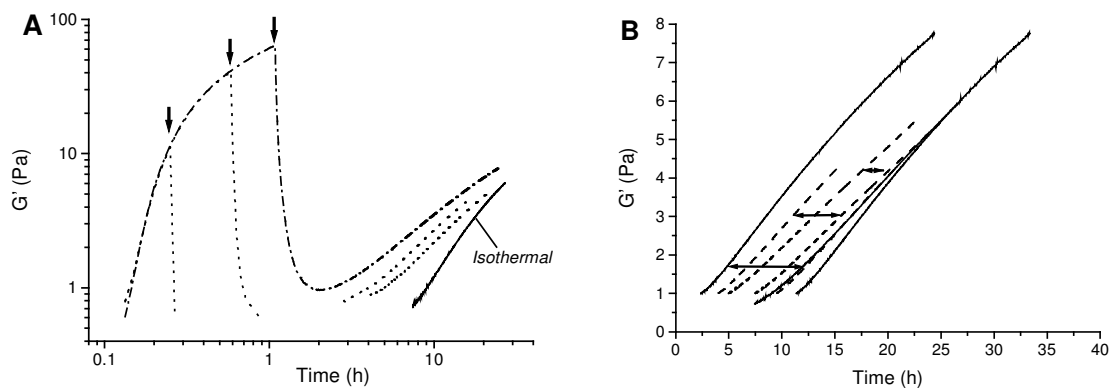

*Figure 7 Memory effect in the close-to-critical regime. **A** shows raw data. One gel (labeled isothermal) was aged at 24°C. Three others were aged for different times at 15°C and then heated to 24°C. Arrows indicate the time at which the gel was heated to 24°C. The heated gels first melt and then re-form at some later time. **B** shows that the data for the cold aged gels after heating can be superposed on the data for the hot aged gel by shifting along the linear time axis. c = 20g/kg*

These gels melted very quickly after heating, but then re-formed some hours later, before a sample aged entirely at 24°C had started to gel. Heating these gels initially melted them, then a wait as long as several hours was needed to discover that they re-gelled. This phenomenon has not been described previously. It occurs because on heating, the system falls below the percolation threshold, but retains some structure. These data illustrate how the spin glass-like dynamics make it hard to define the gel melting temperature in a non-arbitrary way. Figure 4B shows that the kinetics for the cold aged gels can be shifted onto that of the hot aged gel, using linear axes. In fact, the shift factors are linearly proportional to the aging time (data not shown). Some time after heating a cold aged gel, it acts exactly like a gel that has been hot aged for longer. We can call this premature aging. The equivalent experiments in spin glasses can also be superposed by a shift, but the shift is in the opposite direction and so the effect is called rejuvenation [11].

Figure 8 shows what we believe is a new kind of memory effect that occurs after sudden heating in the far-from-critical regime.

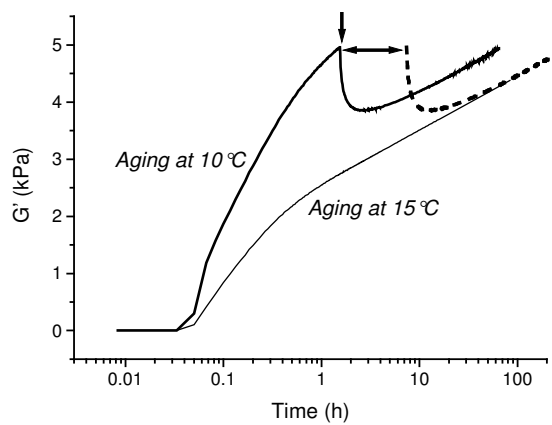

*Figure 8 Memory effect in the far-from-critical regime. The cold aged gel was rapidly heated to 15°C after 1 hour aging at 10°C. Vertical arrow indicates the time of the temperature jump. c = 67g/kg*

The elastic modulus first falls rapidly, but then starts to rise again. In this case a constant gap in log(time) remains. The data for the cold-aged gel can be superposed on that for the hot-aged sample by shifting along the log(time) axis. In this case, the cold aged gel shows accelerated aging. This memory effect has not been observed in spin glasses, either real or simulated.

We have shown that temperature and concentration have equivalent effects on the gelation of gelatin. This same idea has recently become very popular for yield stress fluids and other "jammed" systems [2, 23, 24]. However, the jamming paradigm has only been applied to (apparently) time-independent systems until now. In systems that age, time matters. Scaling the whole time evolution seems a promising approach to understanding, or at least parametrizing, the effects of time on systems jammed far from equilibrium.

Our intuitive picture is that below the critical temperature double helices start to form, but the system quickly become frustrated as all the growing helices are joined by portions of random coil. Such a system quickly arrives at a balance where the energetically favourable formation of helices is exactly balanced by the unfavorable increase in entropic stretch in the attached portions of random coil. The system is then in a marginally stable state. When the temperature is decreased, the stretch decreases, so helix growth starts once again. When the temperature is increased, the coils are over-stretched and undo helices until a new marginal state is reached. This model is clearly very similar to the coupled pendulum model described in Bak's book [25], which is exactly equivalent to the sandpile model [26]. However, this model exhibits self organized criticality and does not age – it reaches a stationary state, so our intuitive picture lacks a vital ingredient.

Sibani and Andersen have recently discussed a model similar to the sandpile model that does age [27]. The extra ingredient that the sandpile model does not have is irreversibility. In gelatine gels, the irreversibility is provided by the co-operativity of the helix-coil transition (see, for instance, [28]). Model spin glasses also show both marginal stability and irreversibility [29]. These shared characteristics of gelatin gels and spin glasses provide some theoretical justification for the similarities in their phenomenologies that we have described here.

The striking similarities shown here between gelatin gels and spin glasses suggest some deep common theory for the dynamics of the two systems. A theory based on hopping over energy barriers - an energy landscape - [1] would be suitable as it is sufficiently specific to make testable predictions, but sufficiently abstract to be applicable to both systems. Studies of gelatin have several clear advantages: gelatin gels have a much longer elementary time scale, so events are observable that are too rapid to observe in real spin glasses. The length scale for gelatin gels is much longer than for spin glasses, so the slow dynamics and fluctuations in the structure can be probed directly using techniques like light scattering. Finally the transparency of the gels and their convenient critical temperature make experiments much easier. We expect more detailed comparisons of spin glasses and gelatin gels to lead to better understanding of universality and specificity in these and other glassy systems.

Acknowledgements
We are very grateful to Jorge Kurchan, Sergio Ciliberto and especially Eric Vincent for helpful advice. A. P. also thanks Erik van der Linden for pointing out the relevance of self-organized criticality and Paolo Sibani for explaining his work.

References
[1] J. P. Bouchaud, L. Cugliandolo, J. Kurchan, and M. Mézard, "Out of equilibrium dynamics in spin-glasses and other glassy systems," in *Spin-glasses and random fields*, A. P. Young, Ed. Singapore: World Scientific, 1998. cond-mat/9702070.
[2] A. J. Liu and S. R. Nagel, "Jamming and rheology: constrained dynamics on microscopic and macroscopic scales," . London and New York: Taylor and Francis, 2001.
[3] P. G. de Gennes, *Scaling concepts in Polymer Physics*: Cornell University Press, 1979.
[4] J. J. Hopfield, "Neural networks and physical systems with emergent collective computational abilities," *Proc. Nat. Acad. Sci. USA*, vol. 79, pp. 2554-2558, 1982.
[5] S. Kirkpatrick, C. D. Gelatt Jr., and M. P. Vecchi, "Optimization by simulated annealing," *Science*, vol. 220, pp. 671-680, 1983.
[6] M. Djabourov, J. Leblond, and P. Papon, "Gelation of aqueous gelatin solutions. II. Rheology of the sol-gel transition," *Journal de Physique*, vol. 49, pp. 333-343, 1988.
[7] G. Tarjus and D. Kivelson, "The viscous slowing down of supercooled liquids and the glass transition: phenomenology, concepts and models," in *Jamming and rheology*, A. J. Liu and S. R. Nagel, Eds. London and New York: Taylor Francis, 2001, pp. 20-38.
[8] V. Normand and J. C. Ravey, "Dynamic study of gelatin gels by creep measurements," *Rheologica Acta*, vol. 36, pp. 610-617, 1997.
[9] W. Kob, "Supercooled liquids and glasses," in *Soft and fragile matter*, M. E. Cates and M. R. Evans, Eds. Bristol: SUSSP and IOP, 2000.
[10] V. Normand, S. Muller, J. C. Ravey, and A. Parker, "Gelation kinetics of gelatin: a master curve and network modelling," *Macromolecules*, vol. 33, pp. 1063-1071, 2000.
[11] K. Jonason, E. Vincent, J. Hammann, J. P. Bouchaud, and P. Nordblad, "Memory and chaos effects in spin glasses," *Phys. Rev. Lett.*, vol. 81, pp. 3243-3246, 1998.
[12] L. Bellon, S. Ciliberto, and C. Laroche, "Memory in the ageing of a polymer glass," *Europhys. Lett.*, vol. 51, pp. 551-556, 2000. cond-mat/9906162.
[13] R. Mathieu, P. E. Jönsson, P. Nordblad, H. A. Katori, and A. Ito, "Memory and chaos in an Ising spin glass," *Phys. Rev. B*, vol. 65, pp. 012411, 2002. cond-mat/0104333.
[14] R. Gilmore, *Catastrophe theory for Scientists and Engineers*. New York: Dover Publications, 1993.
[15] K. Hukushima, H. Yoshino, and H. Takayama, "Numerical study of aging phenomena in short ranged spin glasses," *Prog. Theor. Phys. Suppl.*, vol. 138, pp. 568-573, 2000. cond-mat/9910414.
[16] H. E. Stanley, *Introduction to phase transitions and critical phenomena*. Oxford: Clarendon Press, 1971.

[17] P. G. de Gennes, "Ginsburg parameter and gels," *J. Physique Lett.*, vol. 38, pp. L355, 1977.
[18] M. Adam and D. Lairez, "Sol-gel transition," in *Physical properties of polymeric gels*, J. P. Cohen Addad, Ed. Chichester: John Wiley, 1996, pp. 87-142.
[19] V. Dupuis, E. Vincent, J. P. Bouchaud, J. Hammann, A. Ito, and H. Aruga Katori, "Aging, rejuvenation and memory effects in Ising and Heisenberg spin glasses," *Phys. Rev. B*, vol. 64, pp. 174204, 2001.
[20] J. P. Bouchaud, V. Dupuis, J. Hammann, and E. Vincent, "Separation of time and length scales in spin glasses: Temperature as a microscope," *Phys. Rev. B*, vol. 65, pp. 024439, 2001.
[21] L. Bellon and S. Ciliberto, "Experimental study of the fluctuation-dissipation relation during an aging process," cond-mat/0201224, 2001.
[22] T. Komori, H. Yoshino, and H. Takayama, "Numerical study on aging dynamics in Ising spin-glass models: temperature change protocols," *J. Phys. Soc. Jpn.*, vol. 69, pp. 228-237, 2000. cond-mat/0001395.
[23] A. J. Liu and S. R. Nagel, "Jamming is just not cool any more," *Nature*, vol. 396, pp. 21-22, 1998.
[24] V. Trappe, V. Prasad, L. Cipelletti, P. N. Segre, and D. A. Weitz, "Jamming phase diagram for attractive particles," *Nature*, vol. 411, pp. 772-775, 2001.
[25] P. Bak, *How Nature works*. Oxford: Oxford University Press, 1997.
[26] P. Bak, C. Tang, and K. Wiesenfeld, "Self-organized criticality," *Phys. Rev. A*, vol. 38, pp. 364-374, 1988.
[27] P. Sibani and C. M. Andersen, "Aging and SOC in driven dissipative systems," *Phys. Rev. E*, vol. 64, pp. 021103, 2001.
[28] P. J. Flory, *Statistical Mechanics of chain molecules*. New York: Wiley, 1969.
[29] J. Dall and P. Sibani, "Exploring valleys of aging systems: the spin glass case," cond-mat/0302575, 2003.